\documentclass[journal]{IEEEtran}
\pdfoutput=1
\usepackage[pdftex]{graphicx}
\DeclareGraphicsExtensions{.eps}
\usepackage{amsmath}
\usepackage{color}
\DeclareMathOperator{\dilog}{Li_2}

\newif\ifarxiv
\arxivtrue

\ifarxiv
  \usepackage{draftwatermark}
  \SetWatermarkText{\shortstack{This is the author's version of an article that has been published in this journal. Changes were made to this version by the publisher prior to publication.\\The final version of record is available at
  \quad\quad\quad\quad\quad\color{blue}{http://dx.doi.org/10.1109/LED.2018.2874190}
  \\\\\\\\\\\\\\\\\\\\\\\\\\\\\\\\\\\\\\\\\\\\\\\\\\\\\\\\\\\\\\\\
  \\\\\\\\\\\\\\\\\\\\\\\\\\\\\\\\\\\\\\\\\\\\\\\\\\\\\\\\\\\\\\\\
  \\\\\\\\\\\\\\\\\\\\\\\\\\\\\\\\\\\\\\\\\\\\\\\\\\\\\\\\\\\\\\\\
  \\\\\\\\\\\\\\\\\\\\\\\\\\\\\\\\\\\\\\\\\\\\\\\\\\\\\\\\\\\\\\\\
  \\\\\\\\\\\\\\\\\\\\\\\\\\\\\\\\\\\\\\\\\\\\\\\\\\\\\\\\\\\\\\\\
  \\\\\\\\\\\\\\\\\\\\\\\\\\\\\\\\\\\\\\\\\\\\\\\\\\\\\\\\\\\\\\\\
  \\\\\\\\\\\\\\\\\\\\\\\\\\\\\\\\\\\\\\\\\\\\\\\\\\\\\\\\\\\\\\\\
  \\\\\\\\\\\\\\\\\\\\\\\\\\\\\\\\\\\\\\\\\\\\\\\\\\\\\\\\\\\\\\\\
  Copyright (c) 2018 IEEE. Personal use is permitted. For any other purposes, permission must be obtained from the IEEE by emailing pubs-permissions@ieee.org.
  }}
  \SetWatermarkAngle{0}
  \SetWatermarkColor{red}
  \SetWatermarkLightness{0}
  \SetWatermarkVerCenter{5.5in}
  \SetWatermarkFontSize{7pt}
\fi

\usepackage[normalem]{ulem}
\newcommand{\revisein}[1]{#1}
\newcommand{\reviseout}[1]{}

\begin{document}

\title{Gate-recessed E-mode p-channel HFET with high on-current based on GaN/AlN 2D hole gas}

\author{Samuel~James~Bader,~Reet~Chaudhuri,~Kazuki~Nomoto,~Austin~Hickman,~Zhen~Chen,~%
  Han~Wui~Then, David~A.~Muller,~Huili~Grace~Xing,~\IEEEmembership{Senior Member,~IEEE,}~Debdeep~Jena,~\IEEEmembership{Senior Member,~IEEE}%
\thanks{
  Work was supported by Intel Corporation, AFOSR Grant FA9550-17-1-0048, NSF Grants 1710298, 1534303, and PARADIM (NSF Materials Innovation Platform, DMR-1539918). Work performed at Cornell NanoScale Facility (NNCI member supported by NSF ECCS-1542081), and Cornell Center for Materials Research, supported by NSF MRSEC DMR-1719875.  
HWT with Intel Corporation, other authors with Cornell University. Manuscript received August 26, 2018; revised September 28, 2018. Contact: sjb353@cornell.edu}}

\markboth{IEEE Electron Device Letters,~Vol.~X, No.~Y, August~2018}%
{}
%

\maketitle

\begin{abstract}
  High-performance p-channel transistors are crucial to implementing efficient complementary circuits in wide-bandgap electronics, but progress on such devices has lagged far behind their powerful electron-based counterparts due to the inherent challenges of manipulating holes in wide-gap semiconductors.  Building on recent advances in materials growth, this work sets simultaneous records in both on-current (10 mA/mm) and on-off modulation (four orders) for the GaN/AlN wide-bandgap p-FET structure. A compact analytical pFET model is derived, and the results are benchmarked against the various alternatives in the literature, clarifying the heterostructure trade-offs to enable integrated wide-bandgap CMOS for next-generation compact high-power devices. 
\end{abstract}


%
\IEEEpeerreviewmaketitle
\section{Introduction}
\IEEEPARstart{F}{rom} controlling eco-friendly automotive systems \cite{Su2013, Amano2018, Kachi2014}, to enabling next-generation communications \cite{Yuk2017}, to powering more compact and affordable consumer products \cite{Bindra2015}, major technological shifts place increasingly stringent demands on power and RF electronics.  Gallium Nitride (GaN) is on the forefront of realizing these new applications, given that its large bandgap enables high-power operation, and its built-in polarization can induce dense, undoped, high-mobility electron sheets to provide low on-resistance.  Nonetheless, these advantages have not yet translated to \textit{hole-based} devices in GaN, a deficiency which has severely limited the advance of the technology. Since standard techniques based on complementary p- and n- transistors cannot be straightforwardly integrated in GaN, systems designers must often grit their teeth and slow down their workhorse n-type transistors to interact safely with external driving circuitry \cite{Chu2016}.

This engineering limitation is rooted in the physics of the platform: wide-bandgaps generally lead to heavy valence bands (resulting in lower mobility holes) and deep valence bands, which are difficult to dope and difficult to contact with typical metal workfunctions \cite{Song2010}. For GaN, the only successful chemical dopant is Mg, which has a large activation energy ($.1-.2\ \mathrm{eV} \gg kT$) \cite{Kozodoy2000}, so a high dopant density dominates the electrostatics of a device but provides few free carriers \cite{Nomoto2017}.  The physics of navigating these challenges coincides with massive industrial interest \cite{Amano2018} in advancing power electronics.

The p-doping problem can be addressed as in undoped n-channel devices, by heterostructure design which employs built-in polarization.  High ``polarization-induced doping'' also aids in making contacts \cite{Song2010}, and clever alloy/strain-engineering could mitigate the mobility limitation.  Various authors have produced prototypes based on hole-inducing polar heterostructures (such as GaN/AlN \cite{Li2013}, GaN/AlGaN \cite{Nakajima2014, Nakajima2018, Chu2016, Shatalov2002}, InGaN/GaN \cite{Zimmermann2004, Zhang2016}, or GaN/AlInGaN \cite{Hahn2013, Reuters2014, Hahn2014}). Few of these devices (\cite{Chu2016,Hahn2013,Reuters2014,Hahn2014,Nakajima2018}) have satisfied the circuit designers desire for normally-off (``E-mode'') operation, wherein the device does not conduct without applied gate bias. Among these, the on-currents ($<$10 mA/mm) are generally two orders smaller than in similarly sized n-channel devices.  

The highest on-currents acheived to date are ``normally-on''  devices by the GaN/AlN approach, which maximizes the polarization difference. However, the only reported GaN/AlN E-mode device \cite{Li2013} was produced without gate-specific recess. Consequently, the entire device, not only the gated region, was depleted, such that space-charge-limited transport clipped the device performance.  This work demonstrates the effectiveness of a gate-recess in specifically depleting the GaN/AlN channel to acheive the best performance to-date on this platform.

\section{Experimental Results and Modelling}
\begin{figure}[!t]
\centering
\includegraphics[width=\columnwidth]{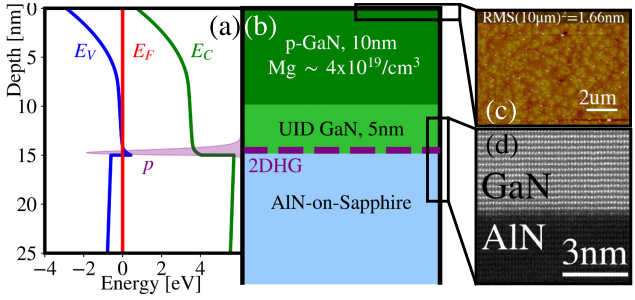}
\caption{(a) Energy-band diagram and (b) layer structure of the grown heterostructure.  Holes (purple shade) are tightly confined to the GaN/AlN interface, forming a 2D carrier gas. (c) AFM scan, showing a relatively rough (RMS 1.66 nm) epi-surface due to the high (for MBE) doping levels, (d) cross-sectional TEM showing an atomically abrupt GaN/AlN interface. }
\label{fig:epi}
\end{figure}

We recently reported \cite{Chaudhuri2018, ChaudhuriIWN2018} the realization of high quality GaN/AlN heterostructures with p-type sheet resistances as low as 7 k$\Omega$/sq, enabled by (1) the enormous hole charge at the binary polarization discontinuity, and (2) the precise interface obtained with Molecular Beam Epitaxy.  The heterostructure of Fig. \ref{fig:epi} \cite{BaderIWN2018} was grown on an AlN-on-Sapphire template, including a 5 nm nominally undoped GaN channel to prevent impurity scattering, followed by a 10 nm heavily Mg-doped p-cap ($N_A \approx 4\times 10^{19}$/cm$^3$) to lower contact resistivity.
A Van der Pauw Hall measurement (by corner In dots) extracted a charge density $\sigma\approx 5.8\cdot 10^{13}$/cm$^2$ and mobility $\mu\approx7.1$ cm$^2$/Vs, for a sheet resistance of $R_{sh}=15\ \mathrm{k}\Omega$/sq. The simulation in Fig \ref{fig:epi}(a), a Poisson/multiband-$k\cdot p$ solution from nextnano \cite{Birner2007}, predicts a hole density of $\approx 5.3\cdot 10^{13}$/cm$^2$.  
Further, we note from simulation that nearly all the holes are confined to the first couple nanometers of the channel at the GaN/AlN interface, so the conduction can be described by a two-dimensional hole gas (2DHG) rather than a volume density.  (Given the deep nature of the Mg acceptor and significant extent of the surface depletion, it is reasonable to expect--for this and similar structures--that integrated hole densities remain below $10^{11}$ holes/cm$^2$ in the p-GaN, more than two orders smaller than the 2DHG density.)

\begin{figure}[!t]
\centering
\includegraphics[width=.475\columnwidth]{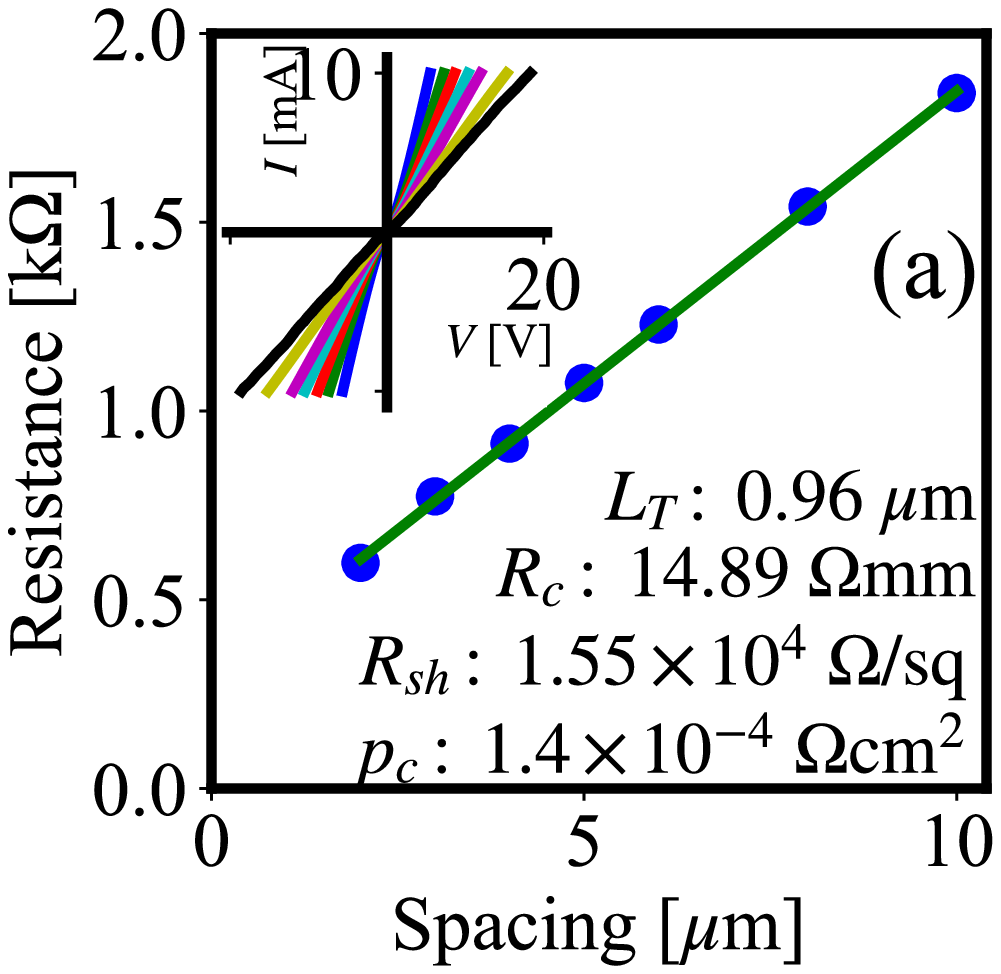}
\includegraphics[width=.51\columnwidth]{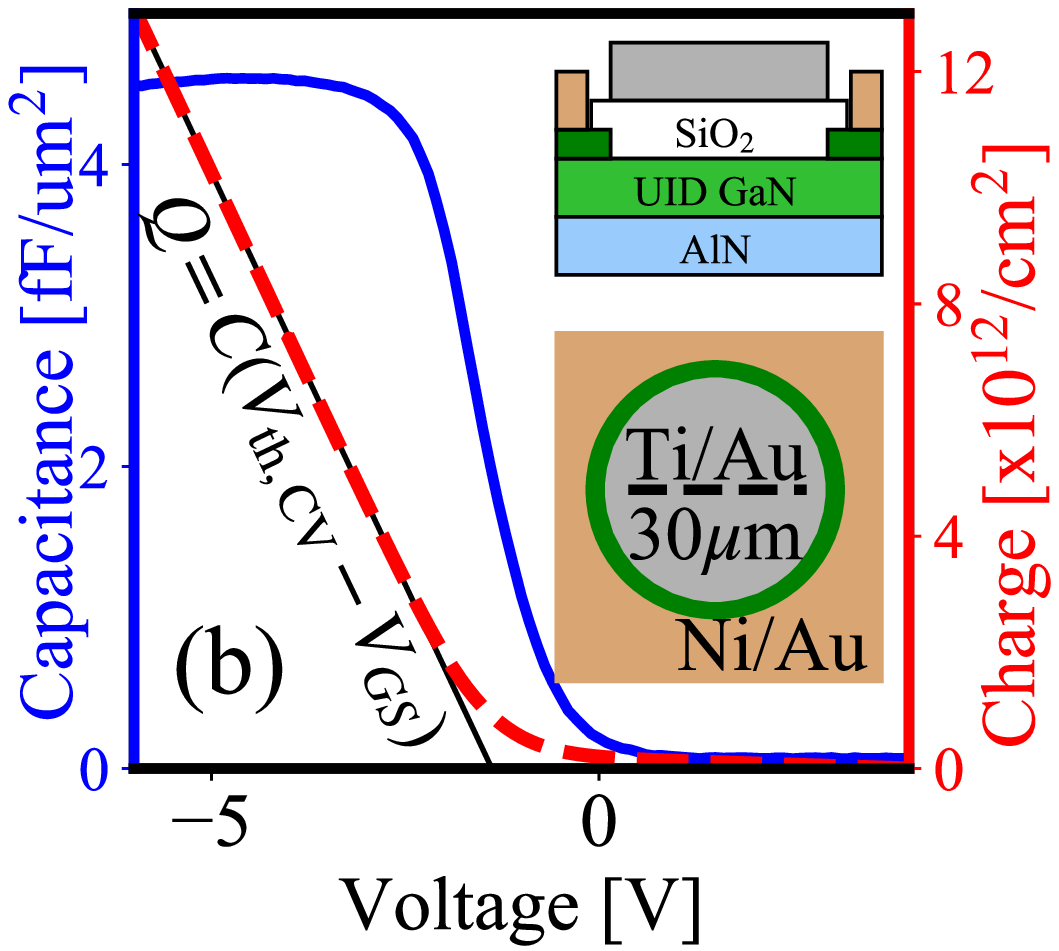}
\caption{Test structures: (a) TLM analysis of Ni/Au ohmic contacts. (b) C-V analysis of a large-area MISCAP.  The capacitance curve (blue) shows classic normally-off p-type 2D behavior, and is integrated to form a charge curve (dashed red), which is described by a linear fit (thin black) beyond threshold.}
\label{fig:test}
\end{figure}

Mesa isolation was performed by a BCl$_3$/Cl$_2$/Ar plasma etch deep into the AlN.  Following hydrochloric and hydroflouric acid cleans, Ni/Au (15/20 nm) contacts were e-beam evaporated and annealed at 450$^\circ$C in O$_2$.  Transfer-length method (TLM) patterns were measured on all dies, demonstrating excellent ohmic contacts to the 2DHG, as analyzed in Fig. \ref{fig:test}(a) for a typical contact resistance, $R_c$, of 15 $\Omega$mm ($\rho_c$ $\sim 10^{-4}\ \Omega$cm$^2$).
Gate recesses 10 nm deep were achieved by a timed, calibrated BCl$_3$ plasma etch and confirmed with AFM.  ALD SiO$_2$ dielectric (target 7 nm) and a Ti/Au gate were deposited.  (As the etch recipe has previously been observed to produce $\sim$45$^\circ$ sidewalls and ALD is highly conformal, step coverage of the dielectric at the recess edge is not a major concern.)  At this point, C-V structures, in Fig \ref{fig:test}(b), showed the signature of a 2D hole gas with a negative threshold and on-state capacitance $C=4.5\ \mathrm{fF}/\mathrm{\mu m}^2$.

\begin{figure*}[!t]
\centering
\includegraphics[width=.75\textwidth]{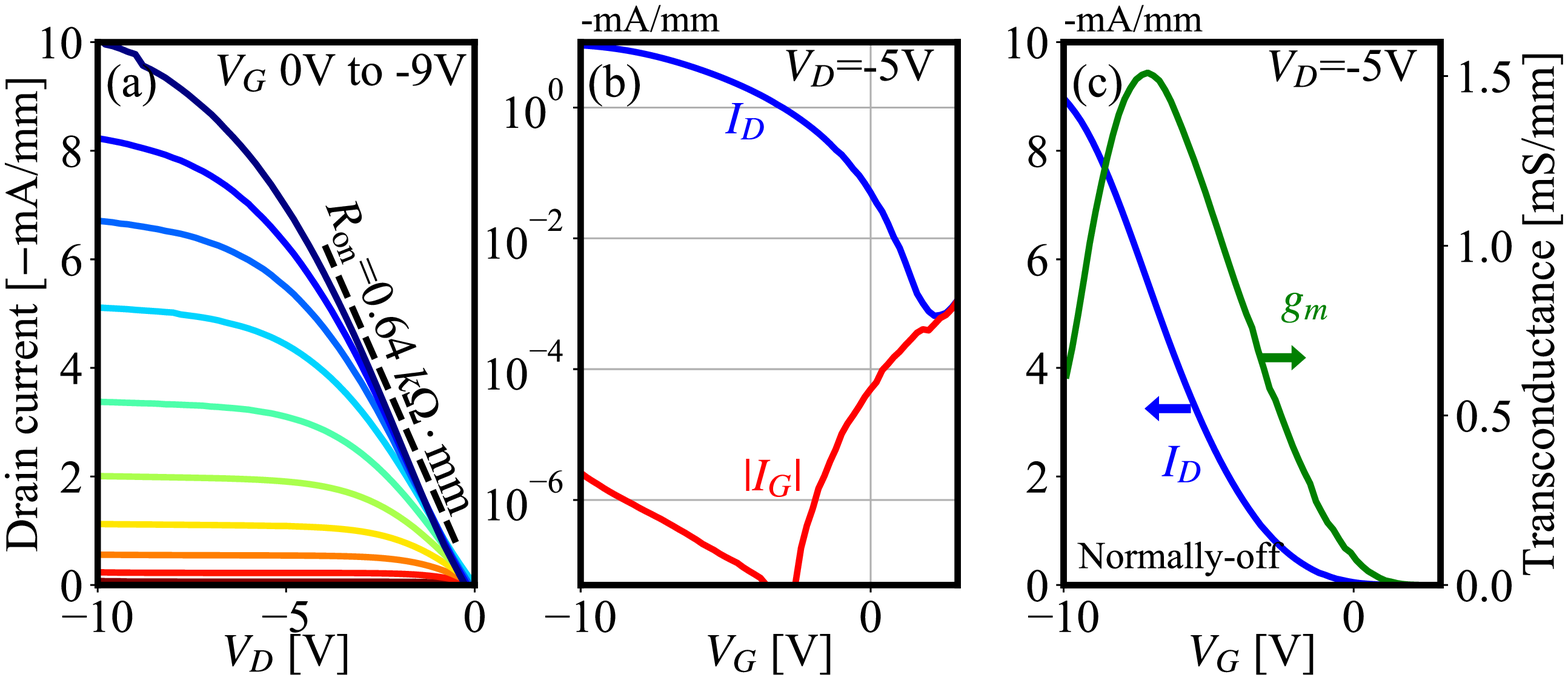}
\includegraphics[width=.23\textwidth]{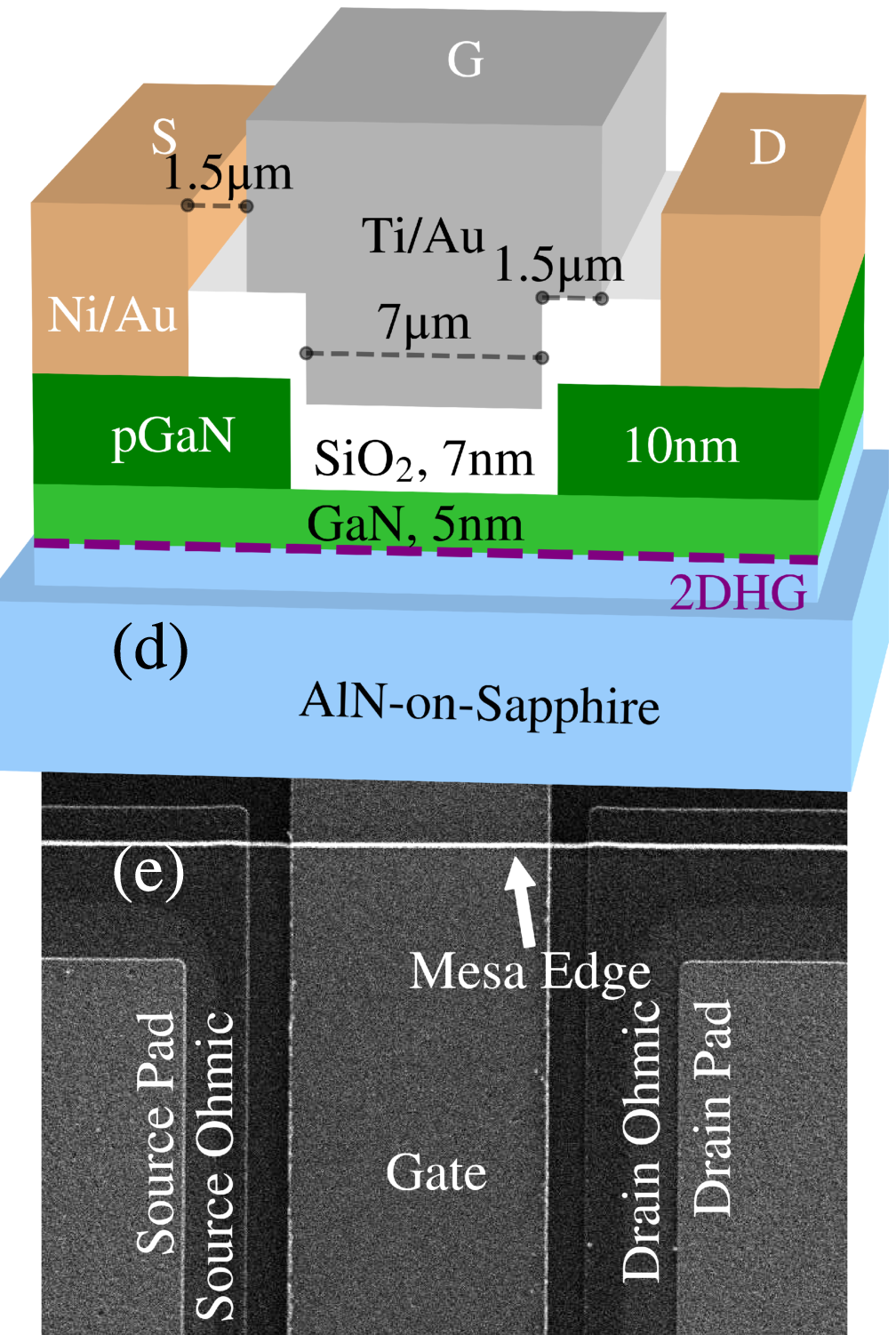}
\caption{Transistor characteristics of a the $L_g=$7 $\mu$m pFET. (a) Output curves show current saturation and an on-resistance of 640 $\Omega\mathrm{mm}$ at $V_G=-9$ V. (b) Log-scale transfer curves show four orders of on-off modulation, limited by gate leakage. Linear-scale transfer curves show normally-off operation and a peak $g_m$ of 1.5 mS/mm.
(d) Cross-sectional schematic of the device structure, indicating the 7um gate length defined by recess through the entire p-GaN layer. The purple dashed line marks the location of the 2D hole gas.  (e) Top-view SEM image of the fabricated pFETs. }
\label{fig:IV}
\end{figure*}

Transistor I-V characteristics, plotted in Fig \ref{fig:IV}, are some of the best reported for wide-gap pFETs.  Clear current saturation and gate control of the wide-bandgap pFET are observed.  Among E-mode GaN/AlN pFETs specifically, the previous record on-current was $4$ mA/mm ($L_g=2\ \mathrm{\mu m}$) but required a drain voltage of $-$40V \cite{Li2013} to be realized.  Due to space-charge-limited transport, that unrecessed device showed less than .04 mA/mm at $V_D=-10$ V.  The $10$ mA/mm shown here (for $L_g=7\ \mathrm{\mu m}$) at $-$10V is not only \textit{more than double} the previous record for the platform \cite{Li2013}, but qualitatively different in that the undepleted access regions pass high currents without demanding enormous drain voltages.  Simultaneously, the $I_\mathrm{on}/I_\mathrm{off}\sim 10^4$ modulation is one order of magnitude larger than that previous flagship device \cite{Li2013}.  Comparing more broadly to the entire selection of E-mode III-Nitride pFETs, it is readily seen that this $10$ mA/mm is already on par with the more-studied, more-scaled E-mode GaN/AlInGaN p-FETs ($\sim$10 mA/mm at $L_g=1\ \mathrm{\mu m}$ \cite{Hahn2013}).  Since the high-quality contacts in this work are not limiting device performance, it is natural that with easily-achievable scaling, these on-currents should show tremendous improvement.

\begin{figure*}[!t]
\centering
\includegraphics[width=\textwidth]{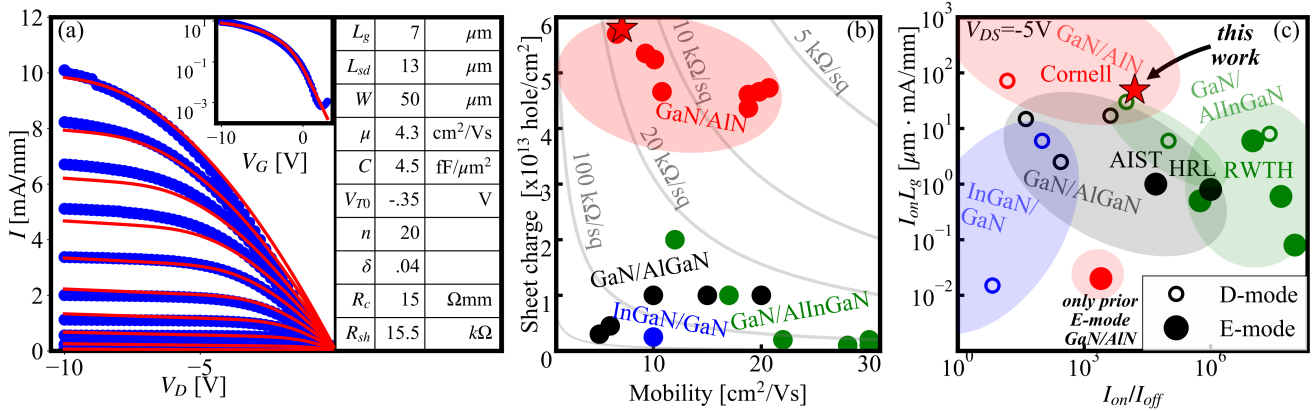}
\caption{(a) Compact model (thin red) fitted to measured data (blue circles). (b) Sheet charge versus mobility for various reported III-Nitride 2D Hole Gases (see also  comparison to other platforms \cite{Chaudhuri2018}).  The GaN/AlN approach has, by a significant margin, the highest sheet charge, due to the extreme polarization discontinuity and its mobility is on par with most other approaches. (c) Benchmark of III-Nitride pFET performance.  The on-current at $V_{DS}=-5V$ is normalized by $L_{g}$ to ignore differences in device scale, since most reports are safely in a long-channel regime.  Among all E-mode devices (filled shapes), this work reports the largest length-normalized on-current thus far, and among all GaN/AlN devices, this work represents the highest modulation.
}
\label{fig:benchmark}
\end{figure*}
The experimental data is modelled as a gradual-channel drift-diffusion FET from a semi-empirical charge-control equation, \cite{Wright1985} $Q=nCV_{th}\ln\left[1+e^{\eta(x)}\right]$ with $\eta(x)=\frac{V(x)-V_{Gi}+V_T}{nV_{th}}$, where $C$ is the gate-channel capacitance, $V_{Gi}$ the intrinsic gate-source voltage, $V_T$ the threshold, $V_{th}$ the thermal voltage, $V(x)$ a local potential, and $n$ the ideality factor.  The channel-integrated current is
\begin{multline}-I_D=\frac{W}{L_g}\mu C (nV_{th})^2 \left(\dilog(-e^{\eta_D})-\dilog(-e^{\eta_S})\right)\\
\mathrm{where}\quad\eta_s=\frac{-V_{Gi}+V_T}{nV_{th}}\quad\eta_d=\frac{V_{Di}-V_{Gi}+V_T}{nV_{th}}\end{multline}
with $\dilog(z)$ the dilogarithm function \cite{Dilog} and $V_{Di}$ the intrinsic drain-source voltage. Access/contact resistances are added to the source and drain as $R_{ext}=R_c+(L_{sd}-L_g)R_{sh}/2$.
Any further drain-induced threshold shift is accounted for by shifting $V_{T}=V_{T0}-\delta V_{Di}$ from its low-bias value $V_{T0}$.

The solid fit of this model to the measurement in Fig \ref{fig:benchmark}(a) points out some interesting details of the device performance.  First, the mobility required to satisfy the model (4.3 cm$^2$/Vs) is lower than that measured by Hall before processing (7.1 cm$^2$/Vs).  There are multiple possible explanations at this point, including plasma damage from the recess etch, increase of the vertical electric field in the gated region, and the difference between Hall-effect and field-effect mobilities.  Further study, such as with gentler digital etching, will be necessary to break apart these effects.  Secondly, the ideality factor is quite large (20), indicating that some mechanism is reducing the efficiency of the depletion.  Body capacitance should be negligible since the underlying material is ultra-widegap aluminum nitride, but the unoptimized dielectric/GaN interface (or perhaps even the GaN/AlN interface) could be contributing traps which reduce the gate efficiency.  Further valence-band-focused dielectric and interface characterization will be vital to maturing this promising pFET platform.

\section{Conclusions and Benchmarking}
Figure \ref{fig:benchmark} (b, c) benchmarks the results against the literature, illuminating several points.  First, among all III-Nitride structures compared in \ref{fig:benchmark}(b), the GaN/AlN approach, powered by the massive binary polarization difference, incorporates the highest hole density, which drastically lowers sheet resistance to the 10 k$\Omega$/sq range to reduce access and contact parasitics. Consequently, see \ref{fig:benchmark}(c), GaN/AlN enables the highest length-normalized on-currents among III-Nitride pFETs.  Additionally, the large bandgaps and band offset enable a thorough pinch-off, with the insulating AlN buffer preventing parasitic n- or p- leakage. 

Other authors have demonstrated basic CMOS inverter operation to varying degrees of success by combining extremely wide p-channel devices with narrow (and relatively low-current) n-channel devices \cite{Chu2016, Nakajima2018, Hahn2014}.  Nevertheless, since the best of the p-channel devices is about two orders of magnitude more resistive than the high-performance n-channel devices to which they may be coupled \cite{Qi2017}, further improvement is essential to making CMOS a serious possibility from a designer perspective.  Toward that end, the device reported here offers obvious avenues for improvement, from basic scaling to gentler digital recess techniques.


In summary, this work has employed gate recess techniques to advance \revisein{the} state-of-art for GaN/AlN wide-gap p-channel devices in both current level and gate control. A suitable compact model was derived, and the device results were benchmarked against the broader III-Nitride pFET literature.  Due to the high sheet conductance and wide bandgaps of the GaN/AlN structure, as well as the opportunity for monolithic integration with powerful AlN/GaN/AlN n-channel devices \cite{Qi2017}, this structure stands out among all competitors as the most promising candidate for inclusion in high-power platforms.
\vfill

{\color{white}
\pagebreak
}
\IEEEtriggeratref{12}

\bibliographystyle{IEEEtran}
\bibliography{IEEEabrv,Pubs-2018EmodepFET}

\end{document}